\documentclass[twoside,leqno,twocolumn]{article}  
\usepackage{ltexpprt} 

\usepackage{graphicx}
\usepackage{subfig}
\usepackage[mathscr]{euscript}
\usepackage{amsmath}
\usepackage{algorithm}
\usepackage{algorithmic}
\usepackage{amsbsy}
\usepackage{bbm}
\usepackage{tabularx}
\usepackage{float}

\newcommand\norm[1]{\left\lVert#1\right\rVert}

\begin{document}

\title{\Large Predicting Multi-actor collaborations using Hypergraphs\thanks{Supported by NSF grant EAGER 1248169.}}
\author{
Ankit Sharma$^{\dag}$
\and 
Jaideep Srivastava$^{\dag}$
\and 
Abhishek Chandra\thanks{Dept. of Computer Science, University of Minnesota, \{ankit,srivasta,chandra\}@cs.umn.edu}}

\maketitle


\begin{abstract} \small\baselineskip=9pt 

Social networks are now ubiquitous and most of them contain interactions involving multiple actors (groups) like author collaborations, teams or emails in an organizations, etc. \(Hypergraphs\) are natural structures to effectively capture multi-actor interactions which conventional dyadic graphs fail to capture. In this work the problem of predicting collaborations is addressed while modeling the collaboration network as a hypergraph network. The problem of predicting future multi-actor collaboration is mapped to hyperedge prediction problem. Given that the higher order edge prediction is an inherently hard problem, in this work we restrict to the task of predicting edges (collaborations) that have already been observed in past. In this work, we propose a novel use of hyperincidence temporal tensors to capture time varying hypergraphs and provides a tensor decomposition based prediction algorithm. We quantitatively compare the performance of the hypergraphs based approach with the conventional dyadic graph based approach. Our hypothesis that hypergraphs preserve the information that simple graphs destroy is corroborated by experiments using author collaboration network from the DBLP dataset. Our results demonstrate the strength of hypergraph based approach to predict higher order collaborations (size$>$4) which is very difficult using dyadic graph based approach. Moreover, while predicting collaborations of size$>$2 hypergraphs in most cases provide better results with an average increase of approx. 45\% in F-Score for different sizes $\in \{3,4,5,6,7\}$ (Figure 6). 

\end{abstract}


\section{Introduction}

The problem of understanding group dynamics is central to the field of Social Sciences. Moreover, the increasing use of internet has led to an exponential increase in amount of online interaction data. As examples, social networking sites like Facebook or Twitter, group communication tools like Skype, Google Hangout, Google Docs, Massive Online multi-player games such as World of Warcraft, etc., are generating social networking data at a massive scale. These social datasets provides minute by minute account of interaction along with the structure and the content of these relationships \cite{Vázquez28122004}. 

In the domain of Social Science, a lot of studies have been conducted to understand how groups form, their static as well as dynamic attributes and structures, and how they evolve over time \cite{coleman1988social}. The research collaborations in scientific community are an excellent example of social networks in which individuals of various expertise collaborate to solve a research problem. Collaboration networks from scientific research community have been extensively used for studying team dynamics \cite{katz1997research}\cite{newman2001structure}\cite{barabasi2002evolution}. Group dynamics has real life applications as well for example in building emergency response teams for natural disasters management, automation of team selection for military operations, etc. 

\begin{figure}[ht]
\begin{minipage}[b]{0.40\linewidth}
\centering
\includegraphics[width=22mm]{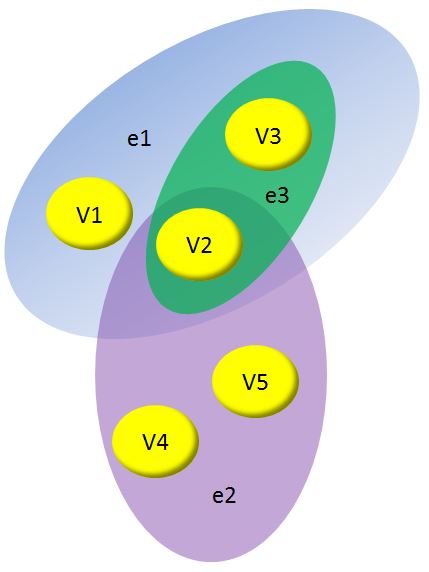}
\caption{Hypergraph}
\label{fig:figure1}
\end{minipage}
\hspace{0.5cm}
\begin{minipage}[b]{0.40\linewidth}
\centering
\includegraphics[width=25mm]{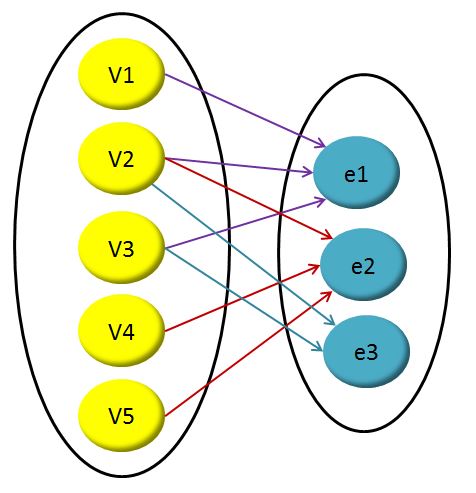}
\caption{Bipartite of hypergraph (Fig.1)}
\label{fig:figure2}
\end{minipage}
\vspace{-1.1em}
\end{figure}

The above examples reveal that there can be multiple overlapping collaborations which form a network of collaborations. Modeling such collaborations in dynamic settings where relationship between actors is evolving over time is a challenging task. 
 Unfortunately, most of the prior research in social network analysis deals with dyadic interactions or small well-defined groups \cite{putnam1996bona} rather than at the group level. There are some studies that have dealt with group interactions by collapsing the group into dyadic links \cite{newman2001structure} and therefore, fail to keep the group level information intact. Ghoshal et. al. \cite{ghoshal2009random} have used tripartite regular hypergraph whcih captures folksonomy data but is too restrictive to capture variable size social collaborations. Guimerà et al. \cite{guimera2005team} attempts attempts to model group using node and group attributes which can explain the network structure but fails to deal with individual group evolution. Bipartite graphs (figure 2) as network models have also been used to capture groups \cite{faust1997centrality} where one set of nodes represent event/collectives/groups and the other second set of nodes represent the actors. But in this model the relations between the actors have to be derived as the group relation is not represented in an intact manner \cite{dhruv2013}. 

Hypergraphs are generalization of graphs, which can have more than two node in an edge (rather than simple graphs where only 2 nodes are part of an edge). Therefore, hypergraphs can easily capture the coexistence of more than two entities in a single relation. Figure 1 shows a hypergraph with five nodes and three edges.

\subsection{Related Work}

Hypergraphs can easily capture the higher-order relationships while incorporating both group and node level attributes. Moreover, research has shown that several social, biological, ecological and technological systems can be better modeled using hypergraphs than using dyadic proxies \cite{estrada2005complex}. There is an abundant literature of hypergraph theory in past \cite{berge1973graphs} and many work in the spectral theory of hypergraphs recently \cite{pearson2012spectral}\cite{xie2013h}. The past decade has also seen an increasing interest for hypergraphs in machine learning community \cite{Zhou07}\cite{tian2009hypergraph}. Hypergraphs have been used to model complex networks in different fields including biology \cite{klamt2009hypergraphs}, databases \cite{fagin1983degrees} and data mining \cite{han1998hypergraph}. In the domain of social sciences, Kapoor et a.l \cite{dhruv2013} have proposed with centrality metrics for weighted hypergraphs. Tramasco et al. \cite{taramasco2010academic} propose hypergraphs based metrics to evaluate various hypothesis, both semantic and structural, regarding team formation. 


Although a lot of work done has been regarding mathematical formulation of hypergraphs, very few works (as stated above) capture the full potential of hypergraph models for real world applications. In this paper, we address the problem of higher order collaboration predictions by modeling it as a hyperedge prediction problem. The collaboration network is modeled as a hypergraph network. Given a previous history of the collaborations we predict collaborations using an supervised approach for hyperedge prediction. In order to capture the time varying hypergraphs and graphs we propose with a novel application of tensor in the form of incidence or hyper-incidence tensors. Our results show that graphs give significantly lower F-Score for higher order groups of size= $\{4,5,6,7\}$ in comparison to hypergraph for most of the cases. In predicting collaborations (hyperedges) higher than size two i.e. more than two entities, hypergraphs in most cases provide better results with an average increase of approx. 45\% in F-Score for different sizes $\in \{3,4,5,6,7\}$ (Figure 6). The main contributions of the paper are summarized as follows:

Our results demonstrate the strength of hypergraph based approach to predict higher order collaborations (size$>$4) which is very difficult using dyadic graph based approach. Moreover, while predicting collaborations of size$>$2 hypergraphs in most cases provide better results with a (25-150)\% increase in F-Score for different sizes $\in \{4,5,6,7\}$  and various training-test splits. 

\begin{itemize}
\item We show a quantitative comparison between graphs and hypergraphs from an applications perspective. This to the best of our knowledge is a pioneer work.
\item We propose a novel application of tensors to capture hypergraphs and use the decomposed tensor's factors to come up with a prediction model for higher order groups.

\item We have successfully addressed the problem of predicting collaborations of higher order which has been rearely dealt with in the past research as it is a problem of considerable complexity.
\end{itemize} 

The rest of the paper is as follows: Section 2 nails down the various hyperedge prediction problems, Section 3 proposes our hypothesis to be evaluated, Section 4 talks about the tensors based algorithm to capture this hypothesis, Section 5 talks about the experiments conducted and results are discussed, which is followed by conclusion and future work.

\section{Hyperedge Prediction problems and Preliminaries}

In this section we describe how higher order collaboration prediction can be mapped to hyperedge prediction problem. 

\subsection{Problem Statement}

In this paper we have used research collaborations where a set of authors ($actors$) collaborate for research. Each of these collaboration results in a $publication$ and each of these publications represents an instance of this collaboration. This means that the same $collaboration$ might result in multiple publications. Each of these $collaboration$ is modeled as a hyperedge in a hypergraph whose each vertex represent an $actor$. The problem of $collaboration$ prediction can then be treated as a problem of predicting a hyperedge. This further splits in two sub-problems: 

\begin{itemize}
\item Problem of predicting the hyperedges already observed in past i.e. \textit{old edge} prediction.
\item Problem of predicting the hyperedges that have never been observed in past i.e. \textit{new edge} prediction. 
\end{itemize} 

To the best of our knowledge these problems have not yet been addressed explicitly. In this paper we are restricting ourselves to the former problem of \textit{old edge} prediction. 

\subsection{Problem Definition}

Let $V = \{v_1,v_2,....,v_{N_a}\}$ be a set of vertices ($actors$). We represent the hypergraph of $collaboration$s using $HG(V,H)$ where $H$ is the incidence matrix of hypergraph which we term as \textit{hyper-incidence matrix}. This matrix \(H\) represent the set of hyperedges ($collaboration$s) $\{h_1,h_2,....,h_{N_h}\}$ where each hyperedge $h_{k} = \{ v_{1}^{h_k},..., v_{|h_k|}^{h_k} \} \subseteq V$. We divide time into small snapshots (of size $w$ as shown in Figure 4) with $t$ as its index. $N_{c}^{t}$ is the number of $publications$ occurred in snapshot $t$ and there are $N_t$ number of snapshots in past. We denote the $i^{th}$ $publication$ in the $t^{th}$ snapshot by \(c_{i}^{(t)}\) \(\forall i = \{1,2,...,N_{c}^{(t)}\}\). Each of this $publication$ $c_i^{(t)}$ represents some $collaboration$ (hyperedge) $h_k$. A mapping function $\phi$ is defined which returns the $collaboration$ (hyperedge) represented by a given $publication$ such that $\phi(c_i^{(t)})=h_k $ $\forall k=\{1,....,N_h\}$ where $N_h$ are the number of distinct $collaborations$ (hyperedges) in the past. Size of $H$ is therefore, $N_h \times N_a$ and we call $s_k$ as the cardinality (No. of vertices inside $h_k$) of the hyperedge $h_k$ i.e. $s_k = |h_k|$.

The problem of \textit{old link} prediction is now defined as follows: Given a past history of collaborations $C_{hist}=\{c^{(t)}\}_{t=1}^{N_t}$ our goal for the problem of \textit{old link} prediction is to predict the likelihood of future occurrence of each of the hyperedges $h_k$ $\forall k=\{1,....,N_h\}$ (i.e. \textit{collaboration}s already observed in past).

\begin{figure}[h!]
\centering
\includegraphics[width=60mm]{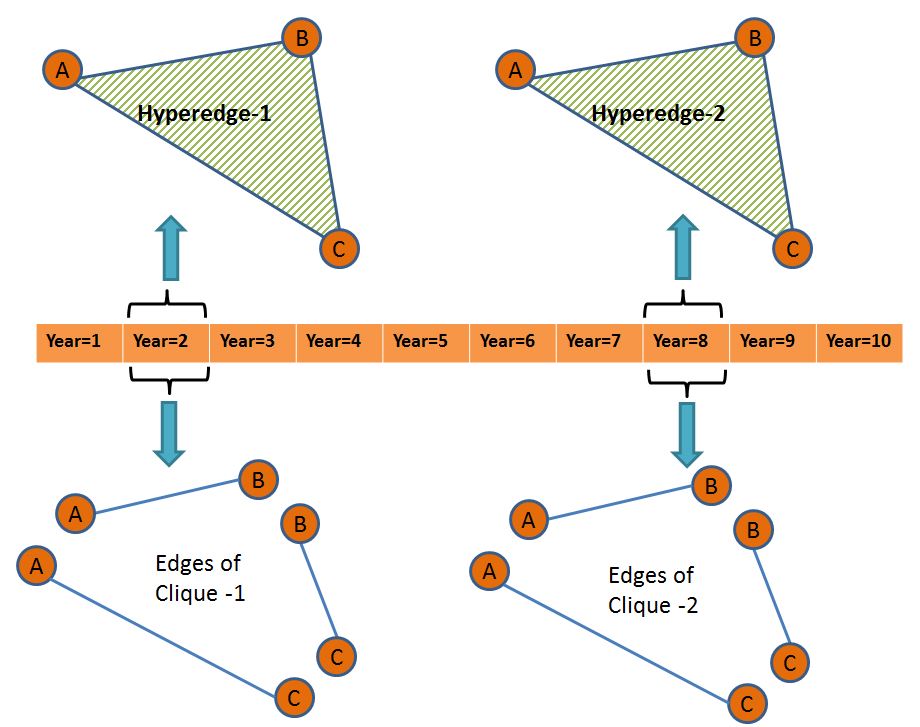}
\caption{Toy Example showing of two $publication$s published by $collaboration$ (A, B, C) in year=2 and year=8 with their hyperedge (top) and clique of dyadic edges (bottom) representation.}
\label{overflow}
\vspace{-1.7em}
\end{figure}

\section{Hypothesis}

In this section we state the hypothesis which is evaluated in this work. We claim that modeling social collaborations or interactions as hypergraph is likely to conserve a lot of information that is destroyed when modeled as dyadic graphs. The claim is supported by the following examples:

\begin{itemize}

\item \textit{Independent dyadic interactions fail to predict higher order interactions :} For example, if A and B talk to each other often and similarly does, the pair (B-C) and (C-A). But this is unable to capture the same information nor can it give a sufficient prediction that (A-B-C) in a group will be interacting together. Whereas if we have seen (A-B-C) together several times this information is completely different than what we can attain from just observing the individual interaction independently. Thus, there is a blatant need for capturing higher order interaction is a form other than dyadic interactions. 

\item \textit{Higher order interactions are captured in a much better manner using hypergraphs than a corresponding dyadic clique based representation:} For example as show in Figure 3, a $collaboration$ of authors A,B and C produced couple of $publications$ in a time window of ten years. Our aim is to predict future likelihood (P(A-B-C)) of this $collaboration$ A-B-C reoccurring. If we use hyperedge representation then P(A-B-C) = 2/10. Whereas, on splitting the hyperedges as cliques of dyadic links, P(A-B-C) = P(A-B)xP(B-C)xP(C-A)= (2/10)x(2/10)x(2/10) = (8/1000) which is clearly less than the probability using the hyperedge. Hypergraph simply keeps the joint probability information intact.
\end{itemize}

\begin{figure}[h!]
\centering
\includegraphics[width=60mm]{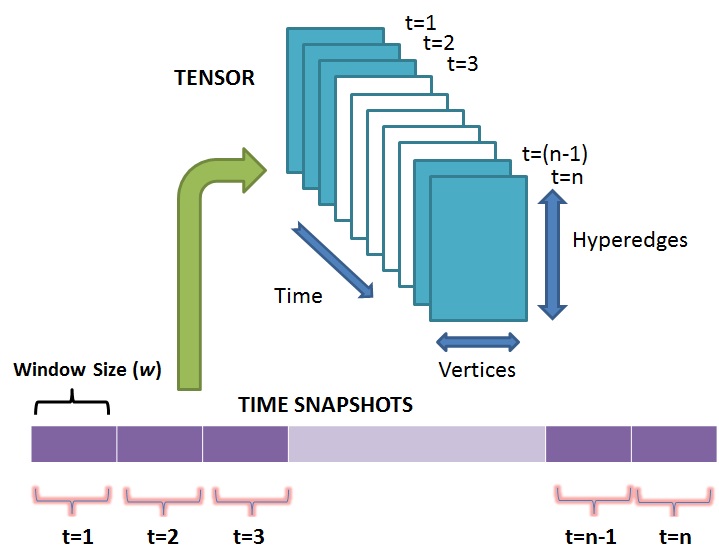}
\caption{A tensor representation of the temporal information (snapshot size=$w$). Each snapshot data is fed in the corresponding hyper-incidence matrix.}
\label{overflow}
\vspace{-1.7em}
\end{figure}

\section{Proposed Approach}

This section describes the approach used to capture the above intuition and build a platform to conduct comparative analysis between graphs and hypergraphs. This section is divided into two sections. In the first section the hypergraph modeling using tensors is explained and the next section described the supervised hyperedge prediction. 

\subsection{Collaboration Modeling}

\subsubsection{Tensor and Incidence matrix representations}

A tensor is a multidimensional, or N-way, array \cite{pearson2012spectral} and has proven to capture multi-dimensional data effectively \cite{kolda07}. For example Tensors allow to handle time as a separate dimension. This provides more flexibility to creatively manipulate the temporal dimension. Moreover, the temporal patterns can be captured using tensors to predict future patterns rather than just immediate future. Recently tensors have already proved effective in predicting temporal link prediction by Dunlavy et al \cite{kolda11}. This has encouraged us to capture hypergraphs and graphs using 3-way tensors where the first two dimensions capture the hypergraph/graph incidence matrix and the third dimension captures the temporal information. Keeping the same incidence matrix representation for both graph and hypergraph allows to have a parity comparison between the two models. 
We denote the tensor for graph and hypergraph using $\mathscr{Z}_{g}$ and $\mathscr{Z}_{h}$ which represent array of the snapshots of incidence or the hyper-incidence matrix respectively (Figure 4). Snapshot \(t\) refers to a time period \(T=(w*(t-1), w*t)\). 

Similar to hypergraph we represent graph as \(G(V,E)\) where the graph incidence matrix is \(E\) represent the set of edges $\{e_1,e_2,....,e_{N_g}\}$. Each edge contains a pair of vertices i.e. \(e_k=\{v_i^{e_k},v_j^{e_k}\} \subseteq V\) (Section 4.1.2 describes the method to obtain these edges). For the snapshot \(t\) we represent incidence matrix for graph as \(E^{(t)}\) and use \(H^{(t)}\) for the hyper-incidence matrix. Here, \(E^{(t)}\) has the dimension (\(N_{g}\times N_{a}\)) where \(N_{g}\) is the number of distinct dyadic edges between any two actors that have been observed uptil current time. Similarly, the dimension of \(H^{(t)}\) is (\(N_{h}\times N_{a}\)) where \(N_{h}\) is the number of distinct multi-actor $collaborations$ (hyperedges) between the actors that are observed till now. Note that in \(E^{(t)}\) and \(H^{(t)}\) only information of publications in snapshot $t$ is stored but they have same dimension for all values of $t$.

\begin{algorithm} 
\caption{PREDICT-COLLAB($C_{hist}$, $isHypergraph$)} 
\label{alg2} 
\begin{algorithmic}[1] 

\STATE $\mathscr{Z}_{h}$ tensor (size \(N_{h}\times N_{a}\times N_{t}\)) initialized to all zeros.

\STATE $\mathscr{Z}_{g}$ tensor (size \(N_{g}\times N_{a}\times N_{t}\)) initialized to all zeros.

\IF{$isHypergraph$} 
\FOR{$c^{(t)} \in C_{hist}$}
\FOR{$c_i^{(t)} \in c^{(t)}$}
\STATE Find $k$ s.t. $\phi(c_i)==h_k$ 
\FOR{$j$ s.t. $v_j \in \{v_{1}^{h_k},..., v_{|h_k|}^{h_k}\}=h_k $} 
\STATE $\mathscr{Z}_{h}(k,j,t) = \mathscr{Z}_{h}(k,j,t) + \frac{1}{s_k}$
\ENDFOR
\ENDFOR 
\ENDFOR 

\STATE $\mathbf{S}_h$ = BUILD-SIMILARITY-MATRIX ($\mathscr{Z}_{h}$, $N_h, N_a$) \\
\RETURN{ \textbf{return} HYPERGRAPH-PROB-VECTOR ($\mathbf{S}_h$, $N_h, N_a$)}

\ELSE
\FOR{$c^{(t)} \in C_{hist}$}
\FOR{$c_i^{(t)} \in c^{(t)}$}
\STATE Find $k$ s.t. $\phi(c_i)==h_k$ 
\STATE $s_k$ is the cardinality of hyperedge $h_k$.
\FOR{Each of the ${s_{k} \choose 2}$ dyadic links, $d_p$ of the hyperedge $h_k$ as a clique.} 
\STATE Find $k'$ s.t. dyadic edge $d_p$ represents the same subset as $c_i$
\FOR{$j$ s.t. $v_j \in \{v_{1}^{d_p},v_{2}^{d_p}\}=d_p $} 
\STATE $\mathscr{Z}_{g}(k',j,t) = \mathscr{Z}_{g}(k',j,t) + \frac{1}{s_k}$
\ENDFOR
\ENDFOR
\ENDFOR
\ENDFOR

\STATE {$\mathbf{S}_g$ = BUILD-SIMILARITY-MATRIX ($\mathscr{Z}_{g}$, $N_g, N_a$)}  \\ 
\RETURN{ \textbf{return} GRAPH-PROB-VECTOR ($\mathbf{S}_g$, $N_g, N_a$)}
\ENDIF
\RETURN
\end{algorithmic}
\end{algorithm}

\begin{algorithm} 
\caption{BUILD-SIMILARITY-MATRIX ($\mathscr{Z}$, $N_a, N_b$) } 
\label{alg2} 
\begin{algorithmic}[1] 

\STATE $\mathbf{S}$ similarity matrix of size $N_a \times N_b$ initialized with all zeros.
\STATE $K$ is the number of components.
\STATE $[\mathbf{\lambda}; \mathbf{A},\mathbf{B},\mathbf{C}]$ = CP-ALS($\mathscr{Z}$)

\FOR{$k \in \{1,2,...,K\}$}
\STATE $\mathbf{S}$ = $\mathbf{S} + \lambda_k \gamma_k \mathbf{a_k} \mathbf{b_k^\top}$
\ENDFOR 

\RETURN{ \textbf{return} $\mathbf{S}$}

\end{algorithmic}
\end{algorithm}

\begin{algorithm} 
\caption{HYPERGRAPH-PROB-VECTOR ($\mathbf{S}_h$, $N_h, N_a$) } 
\label{alg2} 
\begin{algorithmic}[1] 

\STATE $\mathbf{p_h}$ is the probability vector for hyperedge likelihood of length $N_h$ initialized to all one.

\FOR{$i \in \{1,2,...,N_h\}$}
\FOR{$p$  s.t. $v_p \in h_i$}
\STATE $\mathbf{p_h}(i)$ = $\mathbf{p_h}(i) \ast \mathbf{S}_h (i,p)$ 
\ENDFOR 
\ENDFOR 

\RETURN{ \textbf{return} $\mathbf{p_h}$}

\end{algorithmic}
\end{algorithm}

Therefore, $\mathscr{Z}_{g}(:,:,t)=E^{(t)}$ and $\mathscr{Z}_{h}(:,:,t)=H^{(t)}$ both representing array of snapshots of respective incidence matrices. Dimension of $\mathscr{Z}_{g}$ finally becomes \(N_{g}\times N_{a}\times N_{t}\) and $\mathscr{Z}_{h}$ becomes \(N_{h}\times N_{a}\times N_{t}\) dimensional.

\subsubsection{Loading Tensors}

Next step is to extract effective modeling information from historical publication data $C_{hist}$ and feed it into both the graph and hypergraph tensors. The following couple of subsections describe this process for graphs and hypergraphs separately. We are using the following terms interchangeably: hyperedge and collaboration, occurrence of hyperedge and publication; and vertex and actor.

\textbf{Hypergraph Case (Line (3-11) of Algorithm 1):} All hyper-incidence matrices \(H^{(t)}\)  \(\forall t \) have the same dimension and thus, the same number, \(N_{h}\), of unique hyperedges. Each one of these hyperedges \(h_{k}\) \(\forall k \in \{1,2,...,N_{h}\}\) represent a unique collaboration between a subset of actors (vertices) i.e. \(h_{k} \subseteq V\). For each of the publication \(c_{i}^{(t)} \in c^{t}\) for \(i= \{1,2,...,N_{c}^{(t)}\}\) find the \(k \in \{1,2,...,N_{h}\}\) such that \(c_{i}^{(t)}\) represents the same subset of vertices as \(h_{k}\) i.e. $\phi(c_{i}^{(t)})==h_k$. For this index \(k\) of the hyperedge, the tensor is filled as $\mathscr{Z}_{h}(k,j,t)= \frac{m_k}{s_k}$ where \(j\) is the index of each vertex which is the part of the hyperedge \(h_k\), \(s_k\) is the cardinality of the hyperedge \(h_{k}\) and \(m_{k}\) is the multiplicity of the hyperedge \(h_{k}\). 
Multiplicity is calculated as the log (No. of times $h_k$ occurred in $t$), in other words how many times a particular $collaboration$ published some work in snapshot $t$. This process captures the weight of the hyperedge $h_k$ in the hypergraph tensor. The weight of the hyperedge is modeled as \((\frac{m_{k}}{s_{k}})\), as this definition of hyperedge weights is shown to give the best results by Kapoor et al.\cite{dhruv2013}. This whole process is repeated for all the time snapshots. 

\textbf{Graph Case (Line (14-25) of Algorithm 1):} In case of graph also the graph-incidence matrices \(G^{(t)}\) \(\forall t \) have the same dimension and same number, \(N_{g}\), of unique edges. Each of these edges \(g_{k}\) represent a unique set (dyadic collaboration) between two vertices (actors), \(g_{k} =\{v_i^{g_k},v_j^{g_k}\} \subseteq V\). For each publication \(c_{i}^{(t)} \in c^{t}\) for \(i= \{1,2,...,N_{c}^{(t)}\}\) find the $k \in \{1,....,N_{h}\}$ such that $\phi(c_{i}^{(t)})==h_k$. This hyperedge $h_k$ is broken in to ${s_{c} \choose 2}$ dyadic edges and let us denote each of the dyadic link by \(d_{p}\). For each of the \(d_p\) find the index \(k^{'} \in \{1,2,...,N_{g}\}\) for which the \(d_p\) represents the same edge as \(g_{k}\). For this index \(k^{'}\) the tensor is filled as $\mathscr{Z}_{g}(k^{'},j,t)= \frac{m_{k}}{s_{k}}$ where \(j\) is the index of each vertex which is the part of the edge \(d_{p}\), \(s_{k}\) is the cardinality of the hyperedge \(h_{k}\) and \(m_{k}\) is the multiplicity of the hyperedge \(h_{k}\). Thus we model the dyadic link of the clique to get the weight of the original hyperedge \cite{dhruv2013}. Again, this whole process is repeated for all the time snapshots. 

\subsection{Decomposing the tensors (Algorithm 2)}

Next step in the process is to decompose the tensors (loaded in the previous section). These decomposed factors are used in next section for doing $collaboration$ prediction. The method proposed in this paper for decomposition is inspired by CP Scoring using Heuristic (CPH) method of Dunlavy et al. \cite{kolda11} which has already proven successful. This method is based uses the well know CANDECOMP/PARAFAC (CP) \cite{kolda09} tensor decomposition which is analogous to Singular Value Decomposition (SVD) \cite{golub1970singular} and it converts a tensor into sum of rank one tensors. Given a three dimensional tensor $\pmb{\mathscr{X}}$ with size \(J_a \times J_b \times J_c\) its CP decomposition is given by:
\begin{equation} 
\pmb{\mathscr{X}} \approx \sum\limits_{f=1}^F \lambda_{f} \mathbf{a}_{f} \circ \mathbf{b}_{f} \circ \mathbf{c}_{f}
\end{equation} 

where \(\lambda_{f} \in R^{+}\), \(\mathbf{a}_{f} \in R^{J_a}\), \(\mathbf{b}_{f} \in R^{J_b}\), and \(\mathbf{c}_{f} \in R^{J_c}\). Each of the products \(\lambda_{f} \mathbf{a}_{f} \circ \mathbf{b}_{f} \circ \mathbf{c}_{f}\) is called the \(components\) whereas \(\mathbf{a}_{f}\) , \( \mathbf{b}_{f} \) and \(\mathbf{c}_{f}\) are called the \(factors\) of the decomposition. Note that though 
$\norm{\mathbf{a}_f }$=$\norm{\mathbf{b}_f}$=$\norm{\mathbf{c}_f}$=$1$
but these factors are not orthogonal to each other as it is the case in SVD. Also \(\lambda_f\) is the weight for the \(f^{th}\) component. The decomposition is unique, unlike other tensor decomposition methods, resulting in an attractive method for prediction as the factors can be used directly \cite{kolda11}.  Note that matrices $\mathbf{A}$, $\mathbf{B}$ and $\mathbf{C}$ contain the factors $\mathbf{a}_{f}$,$\mathbf{b}_{f}$ and $\mathbf{c}_{f}$ as column vectors. 

\begin{algorithm} 
\caption{GRAPH-PROB-VECTOR ($\mathbf{S}_g$, $N_g, N_a$)} 
\label{alg2} 
\begin{algorithmic}[1] 

\STATE $\mathbf{p_g}$ is the probability vector for edge likelihood of length $N_g$ initialized to all one.

\FOR{$i \in \{1,2,...,N_g\}$}
\STATE $s_i$ is the cardinality of hyperedge $h_i$.
\FOR{Each of the ${s_{i} \choose 2}$ dyadic links $d_p$, of the hyperedge $h_i$ as a clique.}
\FOR{$p$  s.t. $v_p \in d_p$}
\STATE $\mathbf{p_g}(i)$ = $\mathbf{p_g}(i) \ast \mathbf{S}_g (i,p)$ 
\ENDFOR 
\ENDFOR 
\ENDFOR \\

\RETURN{ \textbf{return} $\mathbf{p_g}$}

\end{algorithmic}
\end{algorithm}

Based upon CPH the similarity between the object \(i\) and \(j\) is contained in a similarity matrix $\mathbf{S}$ as the entry at $(i,j)$. This matrix is defined as follows:

\begin{equation}
\mathbf{S} = \sum \limits_{k=1}^{K} \gamma_k \lambda_k \mathbf{a_k} \mathbf{b_k^\top}
\end{equation}

where \begin{equation} \gamma_k = \frac{1}{T_{buf}} \left( \sum \limits_{t=T-T_{buf}+1}^{T} \mathbf{c}_k(t) \right) \end{equation}

$\mathbf{a_kb_k^{\top}}$ for the component $k$ basically represent the similarity between the object pairs in in the $k^{th}$ component. Let the similarity matrix for graph be $\mathbf{S_g}$ (from decomposition of $\mathscr{Z}_{g}$) and for hypergraph be $\mathbf{S_h}$ (from decomposition of $\mathscr{Z}_{h}$). Compression over $T_{buf}$ number of past years (buffer) captures the intuition that only the recent past publications are relevant for prediction. 


\subsection{Predicting Collaborations}

In this step the similarity matrices $\mathbf{S}_g$ and $\mathbf{S}_h$ are used for predicting the edges or hyperedges. Interpretation of the similarity matrix in our approach is as follows. $\mathbf{S}_g(i,j)$ is the likelihood of the $i^{th}$ dyadic edge occurring in future and also contains vertex $j$. Similarly, for case of hypergraph $\mathbf{S}_h(i,j)$ is the likelihood of the $i^{th}$ hyperedge along with vertex $j$ inside it. In short after the tensor decomposition (and the subsequent compression along time dimension) our method outputs a similarity value for all the $actors$ for each $collaboration$ indicating how likely each of these $actors$ can start working with this $collaboration$.

\begin{figure}[h!]
\centering
\includegraphics[width=50mm]{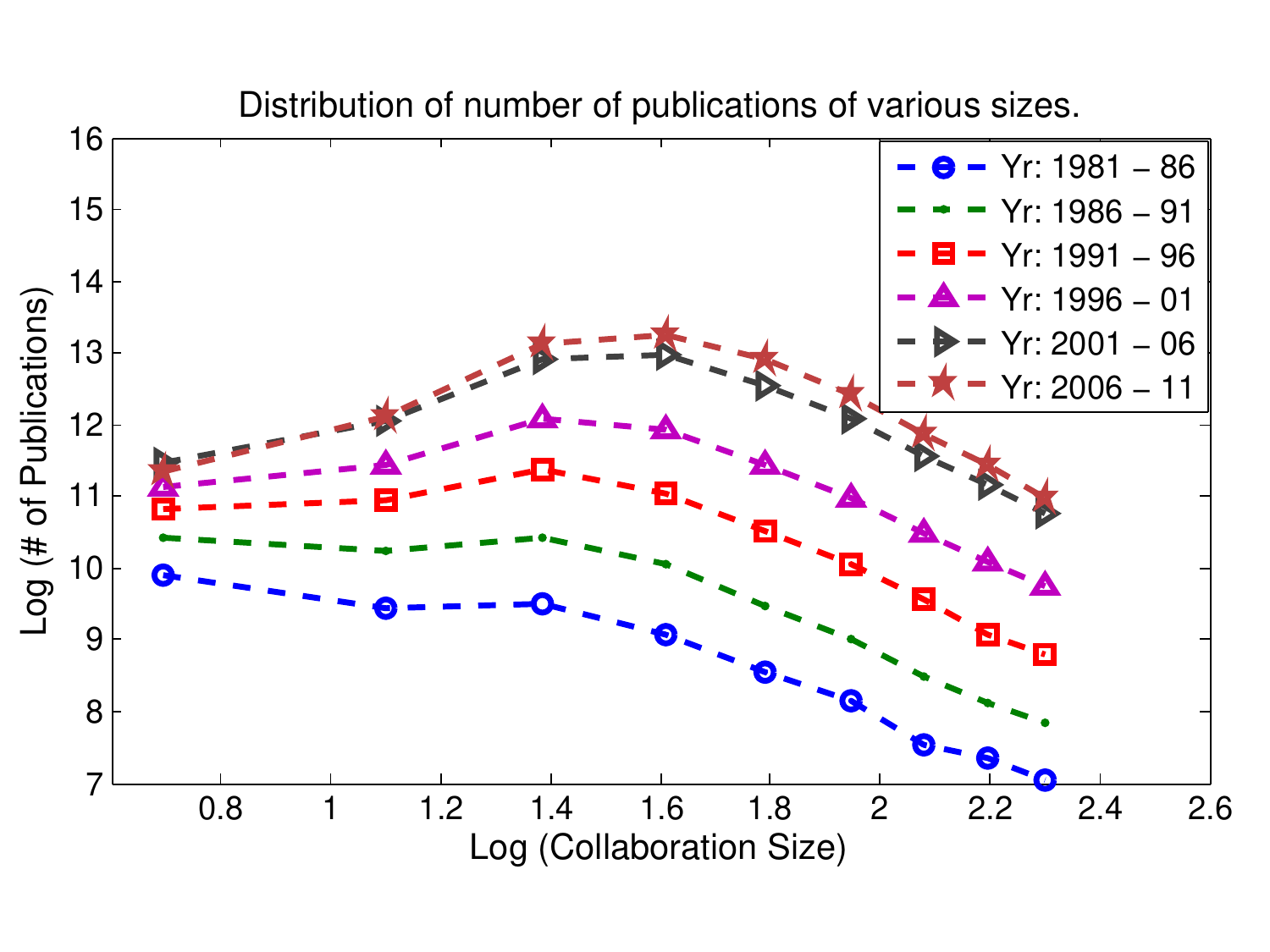}
\caption{Log-Log Plot depicting No. of publications over different sizes of collaboration}
\label{fig:distribution}
\vspace{-1.7em}
\end{figure}

\subsubsection{Hypergraph Case (Algorithm 3):}

If the reassurance of $i^{th}$ hyperedge reoccurs and also contain $j^{th}$ vertex is an event. Assuming that all these events for a particular $i^{th}$ hyperedge for each of the vertices are independent the probability of $i^{th}$ hyperedge reoccurs is defined as: \begin{equation} \mathbf{p_h}(i)=\prod \limits_{p \in h_k} \mathbf{S}_h (i,p) \end{equation} 

\subsubsection{Graph Case (Algorithm 4):}

Similarly, in case of graphs the probability of $i^{th}$ edge reoccurring in future is: \begin{equation} \mathbf{q_g}(i)=\prod \limits_{p \in g_k} \mathbf{S}_g (i,p) \end{equation} The probability of $i^{th}$ hyperedge reoccurring using the dyadic edge probabilities is:  \begin{equation} \mathbf{p_g}(i)=\prod \limits_{q \in D} \mathbf{q_g}(q) =\prod \limits_{q \in D} \prod \limits_{p \in g_k} \mathbf{S}_g (q,p) \end{equation} where $D$ is the set of dyadic edges that are contained in the clique representation of the $i^{th}$ hyperedge.

The outcome of this whole process (Section 4) is these two vectors: $\mathbf{p_g}$ and $\mathbf{p_h}$ . The $i^{th}$ values of $\mathbf{p_g}$ and $\mathbf{p_h}$ are the likelihood of collaboration represented by the $i^{th}$ hyperedge occurring in future as outputted by graph and hyperegraph models respectively. These vectors are used to generate the top-$N$ list as detailed in the Section 5. 

\section{Experimental Analysis}

In this section we discuss the experimental setup used to evaluate the performance of the proposed approach. First section describes the dataset, data preprocessing and experimental setup. In the second section, we discuss the various experiments conducted and their analysis.

\subsection{Dataset and Experimental Setup}

We have evaluated the performance of the proposed approach using the popular DBLP dataset \cite{dblp07} containing publications from years 1930-2011. For the experiments the dataset is divided into training and test periods ($splits$) as shown in the Table 1 and Table 2. As shown in Table 1 the $splits$ are designed with constant training period but variable testing periods. Table 2 contains $splits$ with variable training periods and fixed length testing periods. Table 3 provides the statistics of the training and test set. It provides the total sum of edge counts across all the splits in two different ranges of splits: Split A.1 to A.5 and Split B.1 to B.5 as mentioned. However, only the No. of training and No. of old edges are useful statistics about the data for the proposed experiments.


\begin{figure*}
        \centering
        \includegraphics[width=1.0\linewidth]{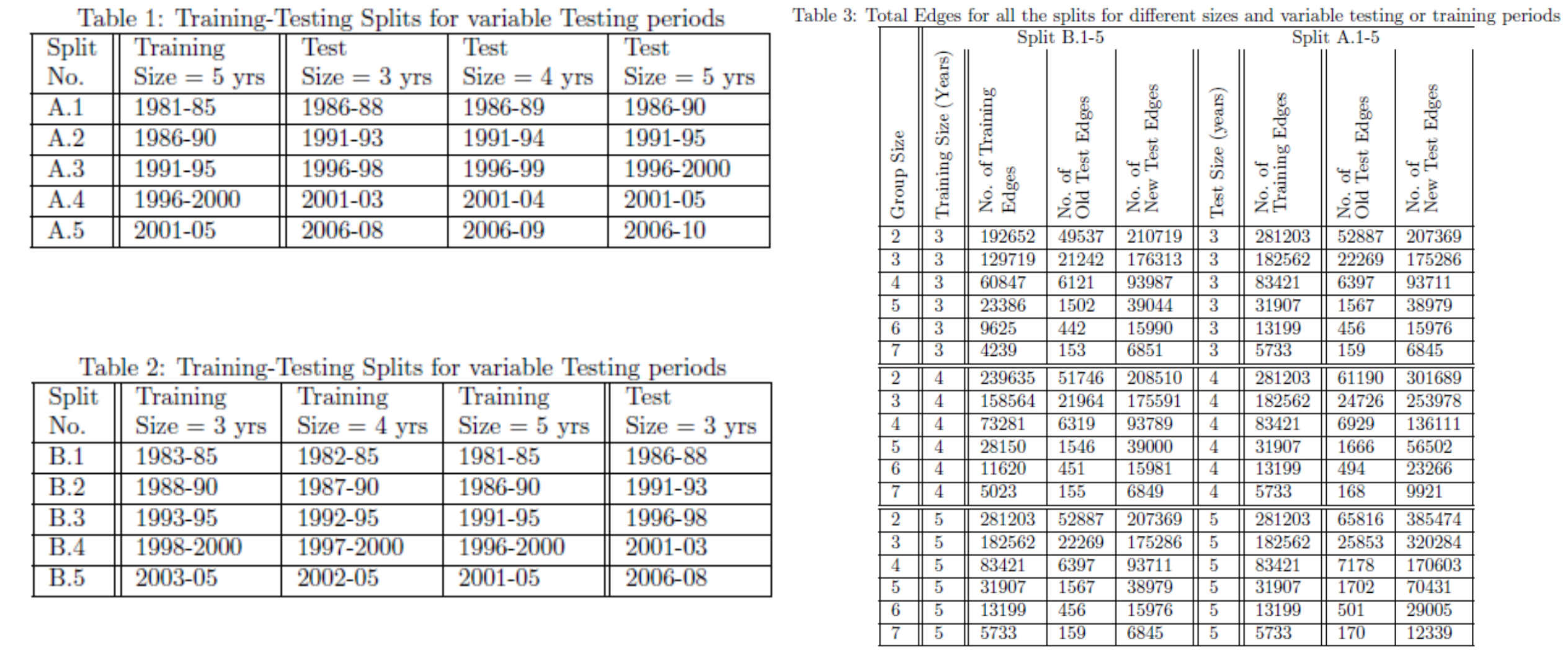}
        \label{fig:TABLES}
\vspace{-1.7em}
\end{figure*}

The distribution Figure~\ref{fig:distribution} is a \textit{log-log} plot showing the distribution of publication counts of various collaboration sizes for different 5 year time periods of DBLP dataset. We observe that the distribution (Figure~\ref{fig:distribution}) across the different intervals follow a similar pattern. This shows that $splits$ that were designed are equivalent as far as conducting experiments is concerned and no bias is involved.

As a preprocessing step, all the single author papers were removed since they do not capture relationships between authors.

For the CP Decomposition (CP-ALS) (that is required for Algorithm 1) Tensor Toolbox \cite{kolda07} is used. To find the parameter $K$ for the CP-ALS algorithm we use the ensemble method approach proposed by Dunlavy et al \cite{kolda11} with $K=\{20,40,...200\}$. Also the parameter $T_{buf}=3$ years is taken \cite{kolda11}. We have used the term graph and dyadic graph interchangeably. 

\begin{figure}
        \centering
        \subfloat{\includegraphics[width=0.245\textwidth]{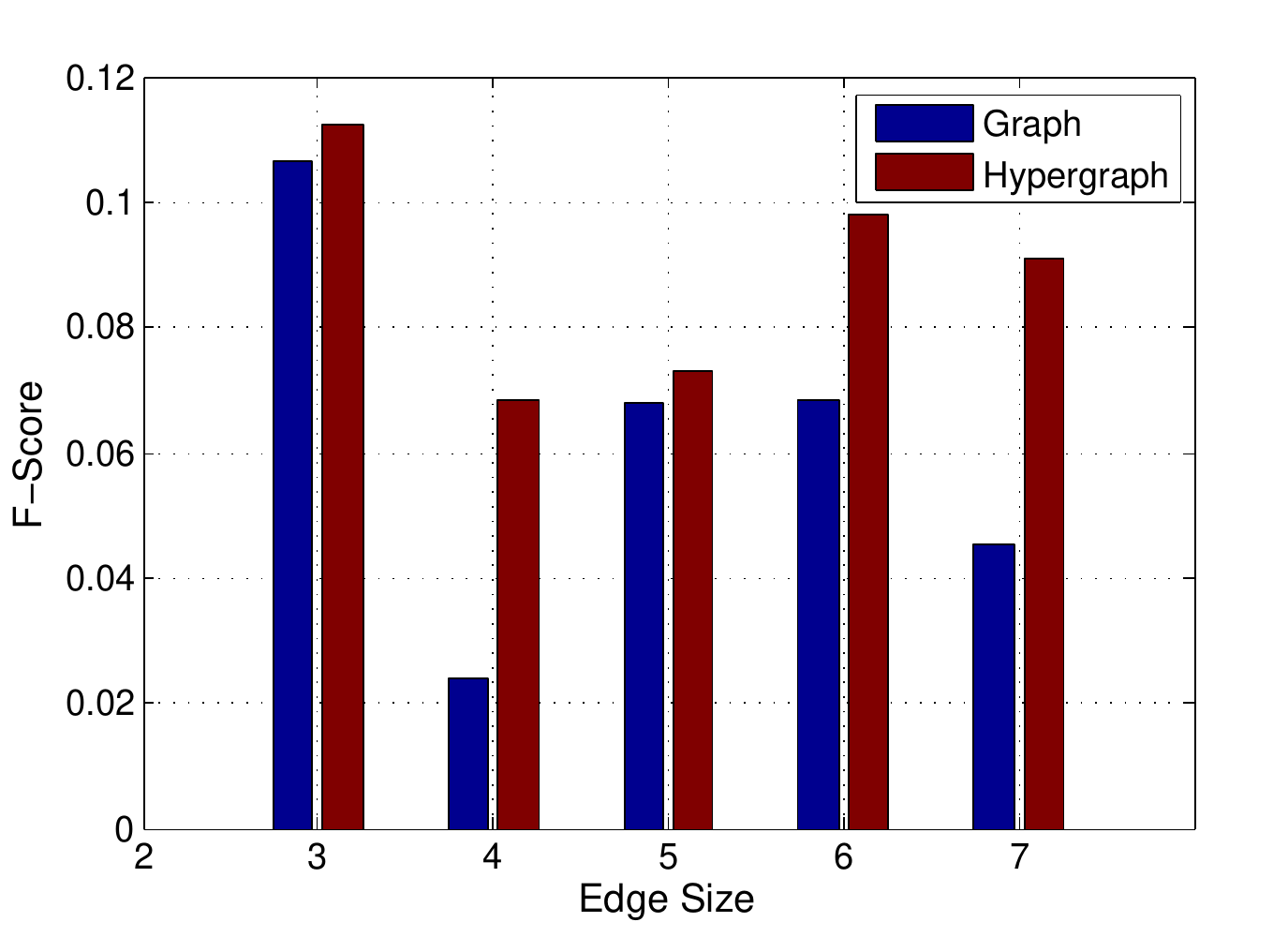}}\subfloat{\includegraphics[width=0.245\textwidth]{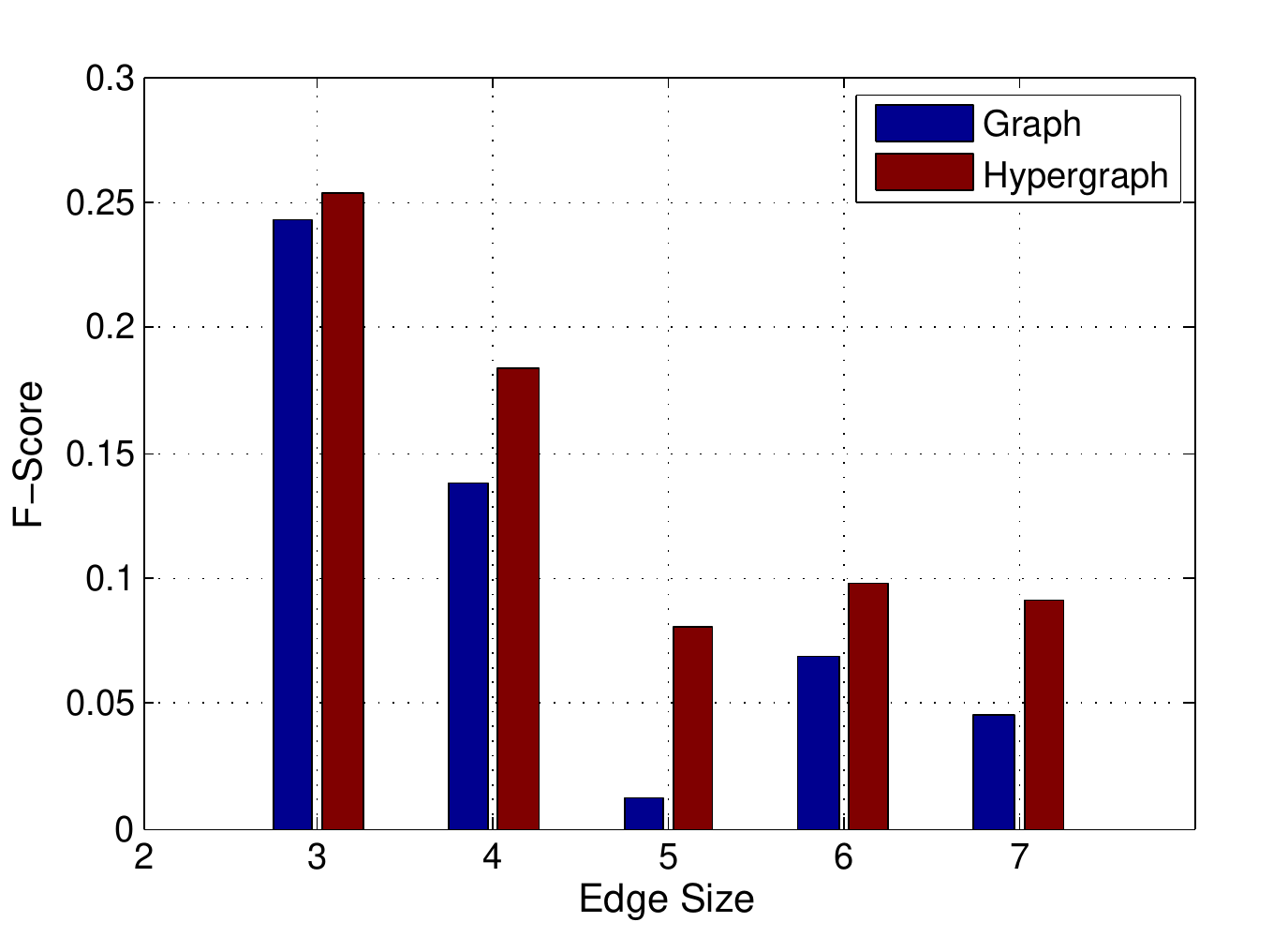}}      
        \caption{Experiment A: (a) AvgF-Score@100 (b) AvgF-Score@1000}
        \label{fig:EXPA}
\vspace{-1.7em}
\end{figure}

\begin{figure}
        \centering
        \subfloat{\includegraphics[width=0.245\textwidth]{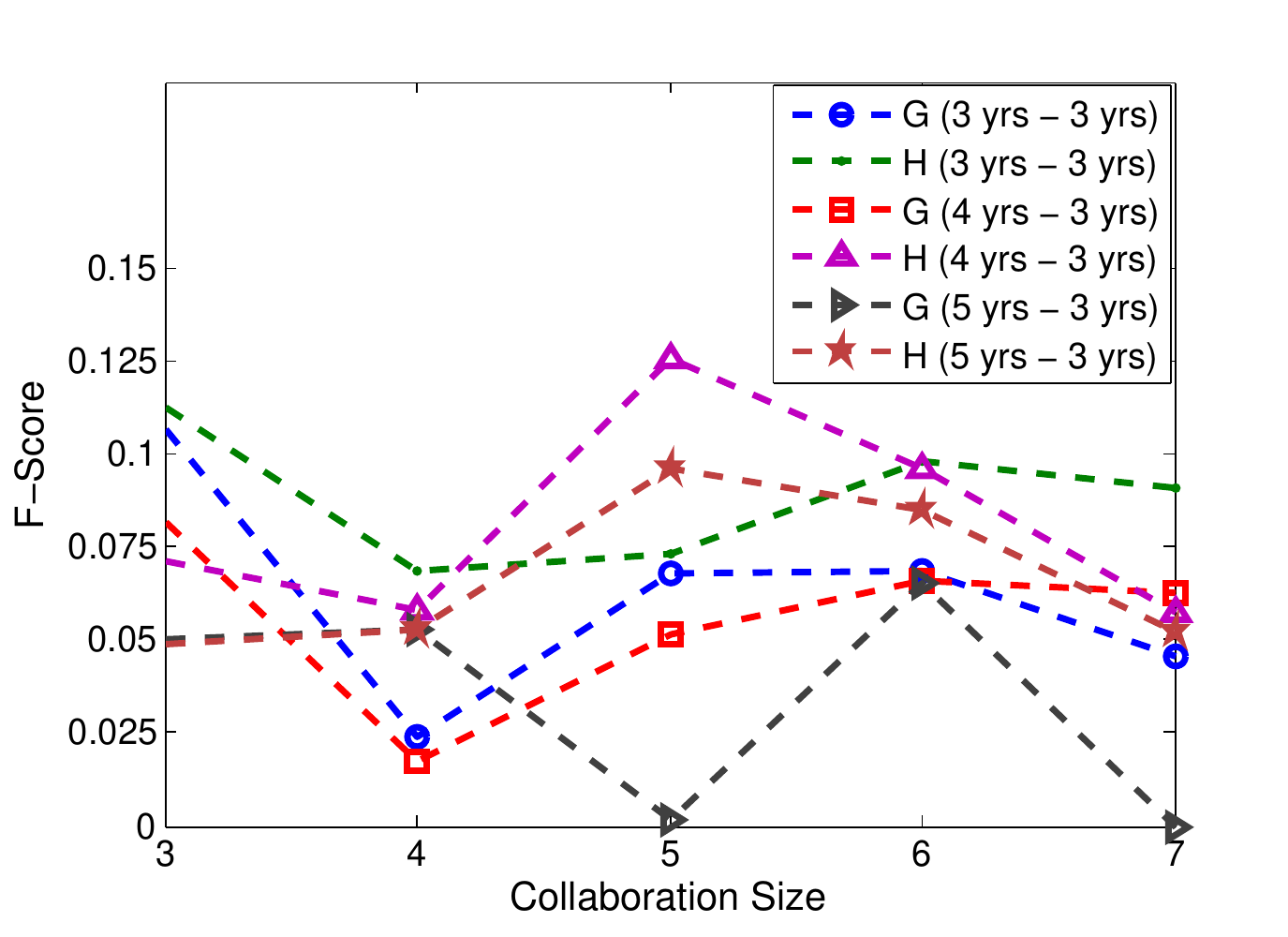}}\subfloat{\includegraphics[width=0.245\textwidth]{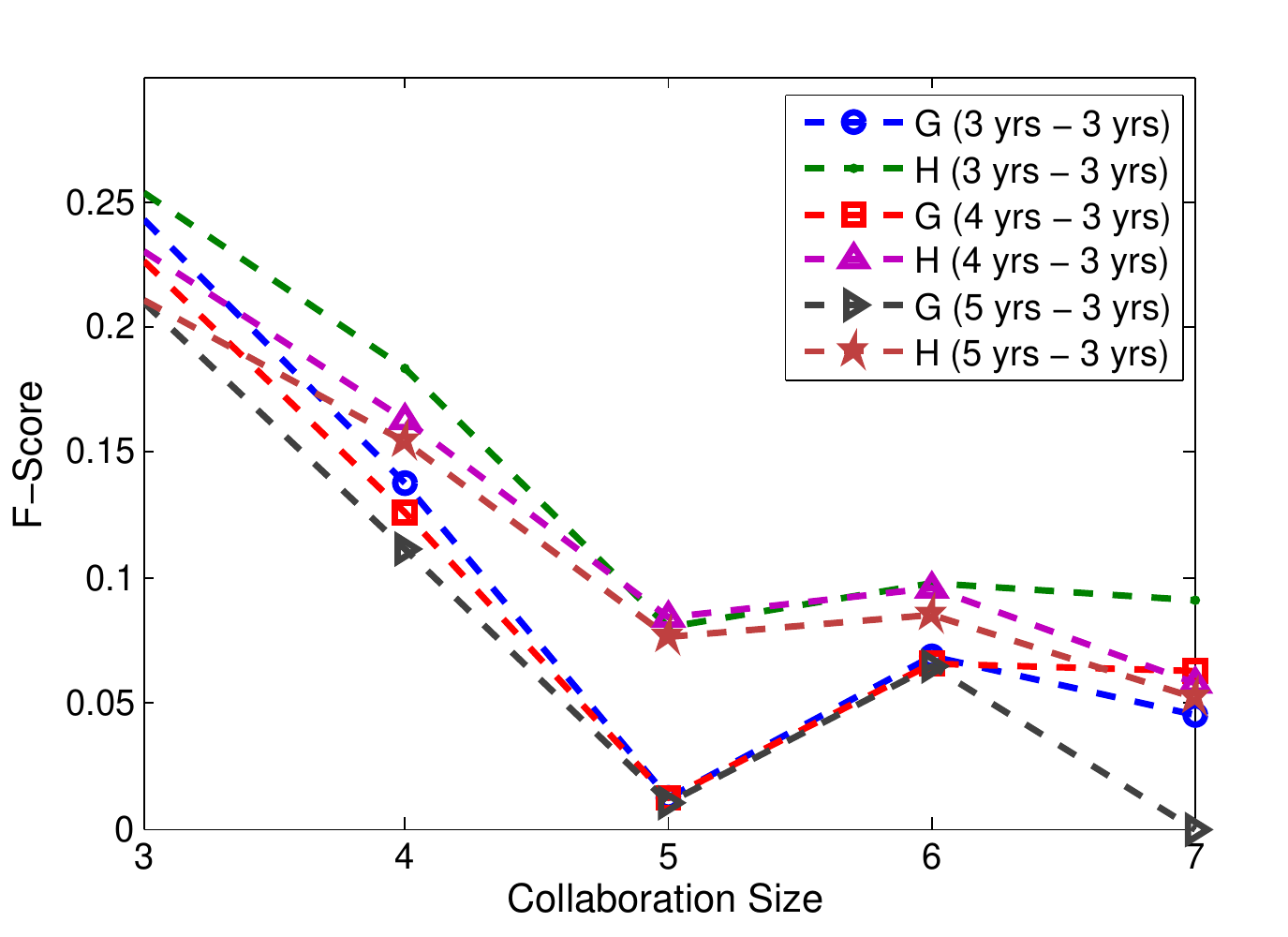}}      
        \caption{Experiment B (Variable Training Size): (a) AvgF-Score@100 (b) AvgF-Score@1000}
        \label{fig:EXPC}
\vspace{-1.6em}
\end{figure}

\subsection{Evaluation} 

In this section four experiments are described that evaluate our proposed approach and provide comparative analysis between dyadic and hypergraphs models. Each of the experiment below is conducted using some of the $splits$. The training period of each $split$ is used to train the dyadic Graph or Hypergraph models using Algorithm 1. The algorithm is run for both graph and hypergraph case to return the edge ($\mathbf{P_g}$) and hyperedge probability ($\mathbf{P_h}$) vectors. These probability vector contains likelihood values for collaborations of different sizes. Each vector is sorted in descending order and the list of top-$N$ elements for each size is extracted. Out of these top-$N$ elements the performance for each size collaboration over test set (old test edges in Table 3) is compared using the following metrics: \\

\begin{equation}
\text{Precision@N (Size-$h$)} = \frac{\substack{\text{\# of size 'h' collaborations} \\  \text{correctly predicted} \\ \text{from size 'h' top-$N$ list}}}{N}\\
\end{equation}


\begin{equation}
\text{Recall@N (Size-$h$)} = \frac{\substack{\text{\# of size 'h' collaborations}\\  \text{correctly predicted} \\ \text{from size 'h' top-$N$ list}}}{\substack{\text{\# of actual size 'h'}\\ \text{collaborations}}}\\
\end{equation}

\begin{equation}
\text{AvgPrecision@N (Size-$h$)} = \frac{\substack{\text{Sum of Precision@N (Size-$h$)} \\  \text{for all splits}}}
					{\substack{{\text{Total \# of splits}}}}\\
\end{equation}

\begin{equation}
\text{AvgRecall@N (Size-$h$)} = \frac{\substack{\text{Sum of Recall@N (Size-$h$)} \\  \text{for all splits}}}
					{\substack{{\text{Total \# of splits}}}}\\
\end{equation}

\begin{equation}
\text{AvgF-Score@N (Size-$h$)} = \frac{\substack{\text{2 * AvgPrecision@N (Size-$h$)} \\ \text{* AvgRecall@N (Size-$h$)} }}  
					{\substack{\text{AvgPrecision@N (Size-$h$)} \\ \text{+ AvgRecall@N (Size-$h$)} }}\\
\end{equation}

This study considers collaborations of size = $\{2,3,4,5,6,7\}$ as we are interested only in higher order collaborations (size=2 is used in the analysis as the trivial dyadic case). AverageF-Score@N and AverageF-Score@N are used in the experiments only for collaborations of size = $\{3,4,5\}$ across all the $splits$ (over which the experiment is conducted). Collaboration of size = $\{6,7\}$ the number of predictions are quiet less as compared to size = $\{3,4,5\}$ case. Therefore, for these cases all the predictions (rather than top-$N$) are used using the following metrics:  

\begin{equation}
\text{Precision (Size-$h$)} = \frac{\substack{\text{\# of size 'h' collaborations} \\  \text{correctly predicted}}}
					{\substack{\text{Total \# of size 'h'} \\ \text{ predicted}}}\\
\end{equation}

\begin{equation}
\text{Recall (Size-$h$)} = \frac{\substack{\text{\# of size 'h' collaborations} \\  \text{correctly predicted}}}
					{\substack{\text{\# of actual size 'h'} \\ \text{collaborations}}}\\
\end{equation}

\begin{equation}
\text{AvgPrecision (Size-$h$)} = \frac{\substack{\text{Sum of Precision (Size-$h$)} \\  \text{for all splits}}}
					{\substack{{\text{Total \# of splits}}}}\\
\end{equation}

\begin{equation}
\text{AvgRecall (Size-$h$)} = \frac{\substack{\text{Sum of Recall (Size-$h$)} \\  \text{for all splits}}}
					{\substack{{\text{Total \# of splits}}}}\\
\end{equation}

\begin{equation}
\text{AvgF-Score (Size-$h$)} = \frac{\substack{\text{2 * AvgPrecision (Size-$h$)} \\ \text{* AvgRecall (Size-$h$)} }}  
					{\substack{\text{AvgPrecision (Size-$h$)} \\ \text{+ AvgRecall (Size-$h$)} }}\\
\end{equation}

In the expriments below AverageF-Score (Size-$h$) is used as a metric to evaluate the collaboration of size = $\{6,7\}$ across all the $splits$ (over which the experiment is conducted).

\subsubsection{Experiment A}

This experiment is conducted over the splits A.1 to A.5 for a fixed test period of 3 years (i.e. from Table 1 column 2 are the training periods and column 3 are the corresponding testing periods). AvgF-Score@100 and AvgF-Score@1000 are show in the Figure 6 (a),(b) for size $= \{3,4,5\}$. As shown in Figure 6(a),(b) for size$=3$, graphs perform comparably with hypergraphs however for size$=4$ prediction using hypergraphs show approx. 150\% and 40\% increase in F-Score for @100 and @1000 cases respectively. For size$>$5 Figure 6(a) and Figure 6(b) are identical showing the AvgF-Score. For size$>$5 graphs show similar trends as for size$=4$ with performance degrading as size increases. As shown in figure 6(a) the F-Score for hypergraph perform better with an increase ranging from 25\% for size$=5$ to almost 100\% for size$=7$. This indicates that Hypergraphs maintain the higher order group information intact. However, owing to the limited training set for higher order (size$>6$) collaborations hypergraph performance is reduced. 

\subsubsection{Experiment B}

This experiment compares the prediction power of the two models: graph and hypergraphs, when trained over variable size training periods. The time period splits used in this case are B.1 to B.4 (which has fixed test period of 3 years) over training period size from 3 to 5 years as show in the Table 2. For size$=\{3,4,5\}$ AvgF-Score@100/1000 and AveragrF-Score for size$>5$ curves for different training periods are shown in Figure 7(a),(b). As shown in Fig 7(a),(b) the F-Score curves for graph model are always lower than hypergraph curves for all size collaborations. Another interesting thing to note is that green curves of hypergraph are above pink and pink is above maroon for most sizes in both Figure 7(a),(b). Similar case is there for graphs (blue above red and red above grey). Thus, increasing the training period in several cases results in decrease in prediction power for both graphs and hypergraphs. This shows that the information about past can act as a noise and thus, decrease prediction accuracy.

\subsubsection{Experiment C}

\begin{figure}
        \centering
        \subfloat{\includegraphics[width=0.245\textwidth]{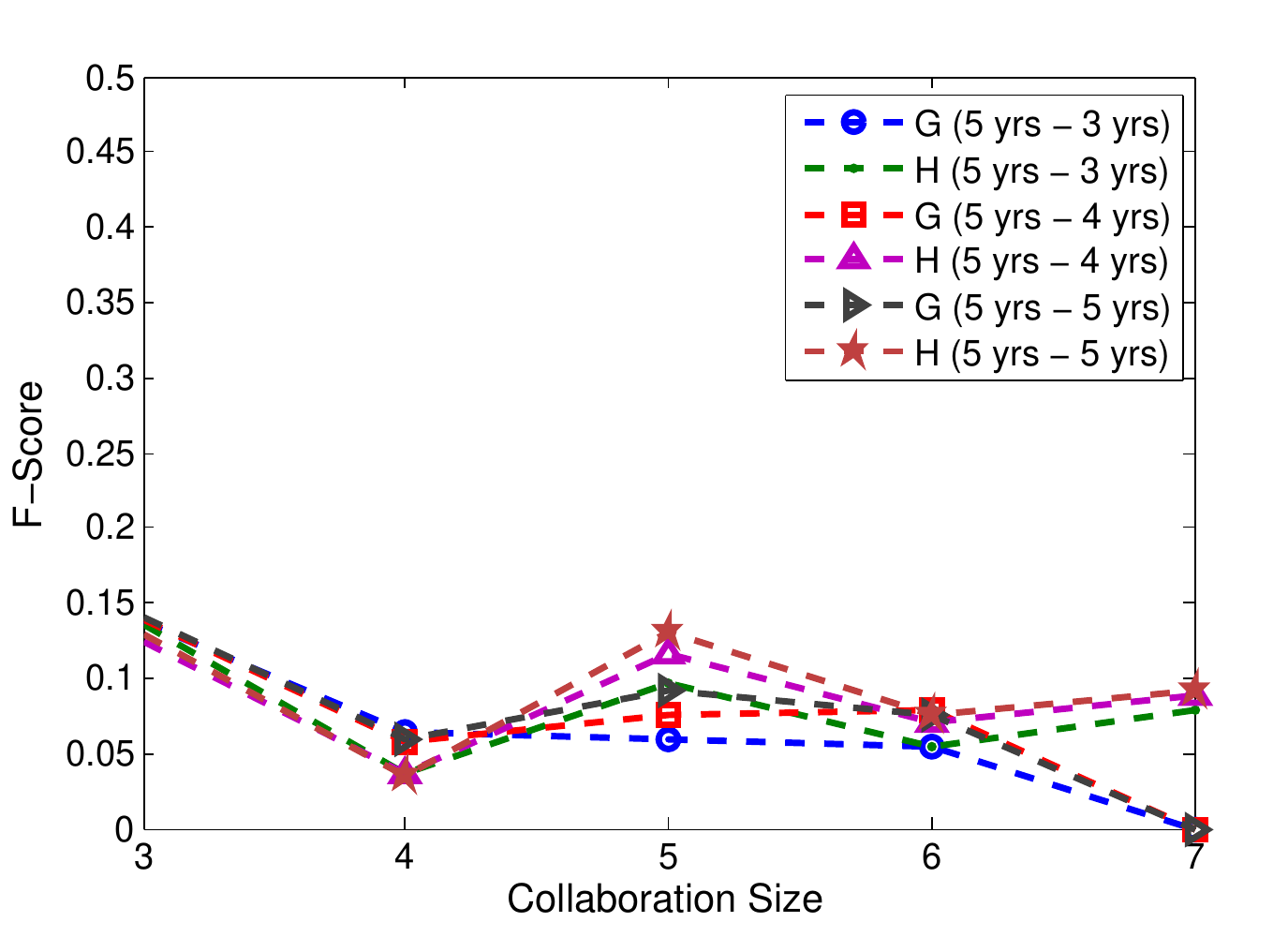}}\subfloat{\includegraphics[width=0.245\textwidth]{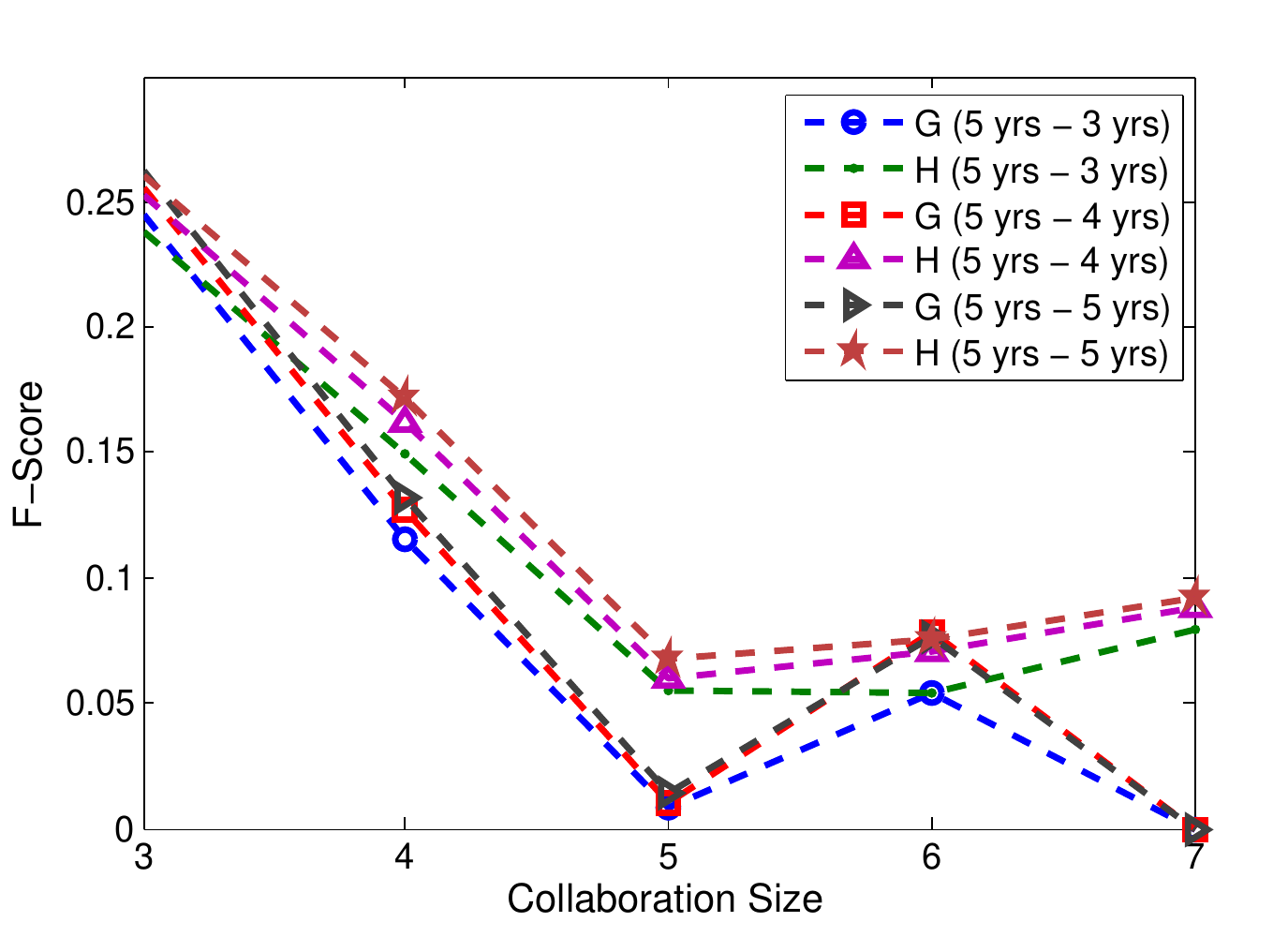}}      
        \caption{Experiment C (Variable Test Size): (a) AvgF-Score@100 (b) AvgF-Score@1000}
        \label{fig:EXPB}
\vspace{-1.6em}
\end{figure}

To get further confidence in the prediction power of hypergraphs we ran experiments with predictions over variable testing periods from three to five years using the splits A.1 to A.4 (Table 1) and fixed training period size$=5$ years. For size = $\{3,4,5\}$ the AvgF-Score@100 and AvgF-Scorel@1000 curves for different testing periods are shown in Figure 8(a),(b). As shown in Figure 8(a),(b), for size = $\{3,4\}$ the graph model (curves colored blue, red and gray) is comparable to the green, pink and maroon curves (which represent hypergraph). However at higher order collaborations hypergraph outperform graphs (as inferred from the AvgF-Scores for size $>=5$ shown in Figure 8(a),(b)).



%

\subsubsection{Experiment D}

\begin{figure}
        \centering
        \subfloat{\includegraphics[width=0.245\textwidth]{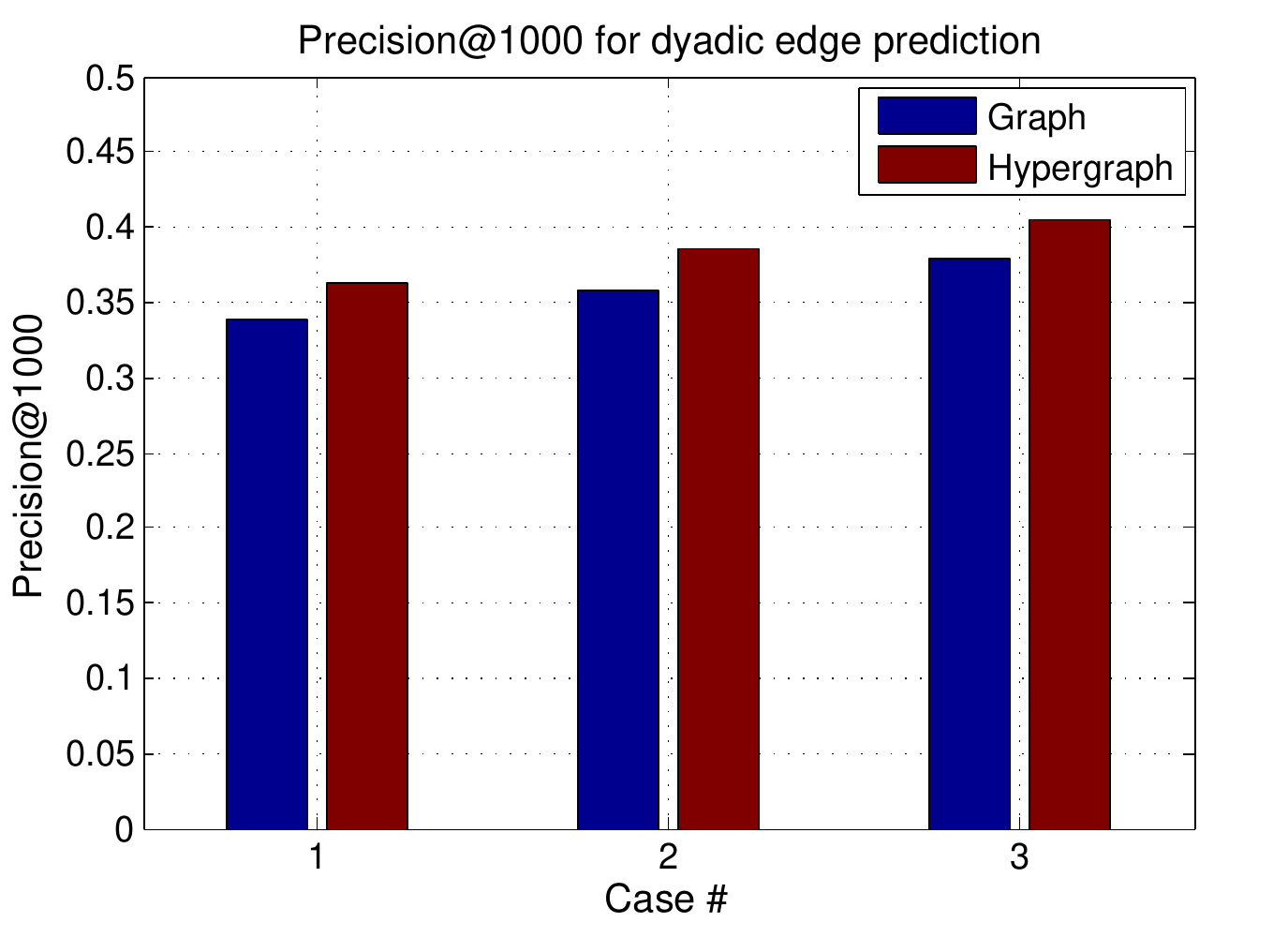}}\subfloat{\includegraphics[width=0.245\textwidth]{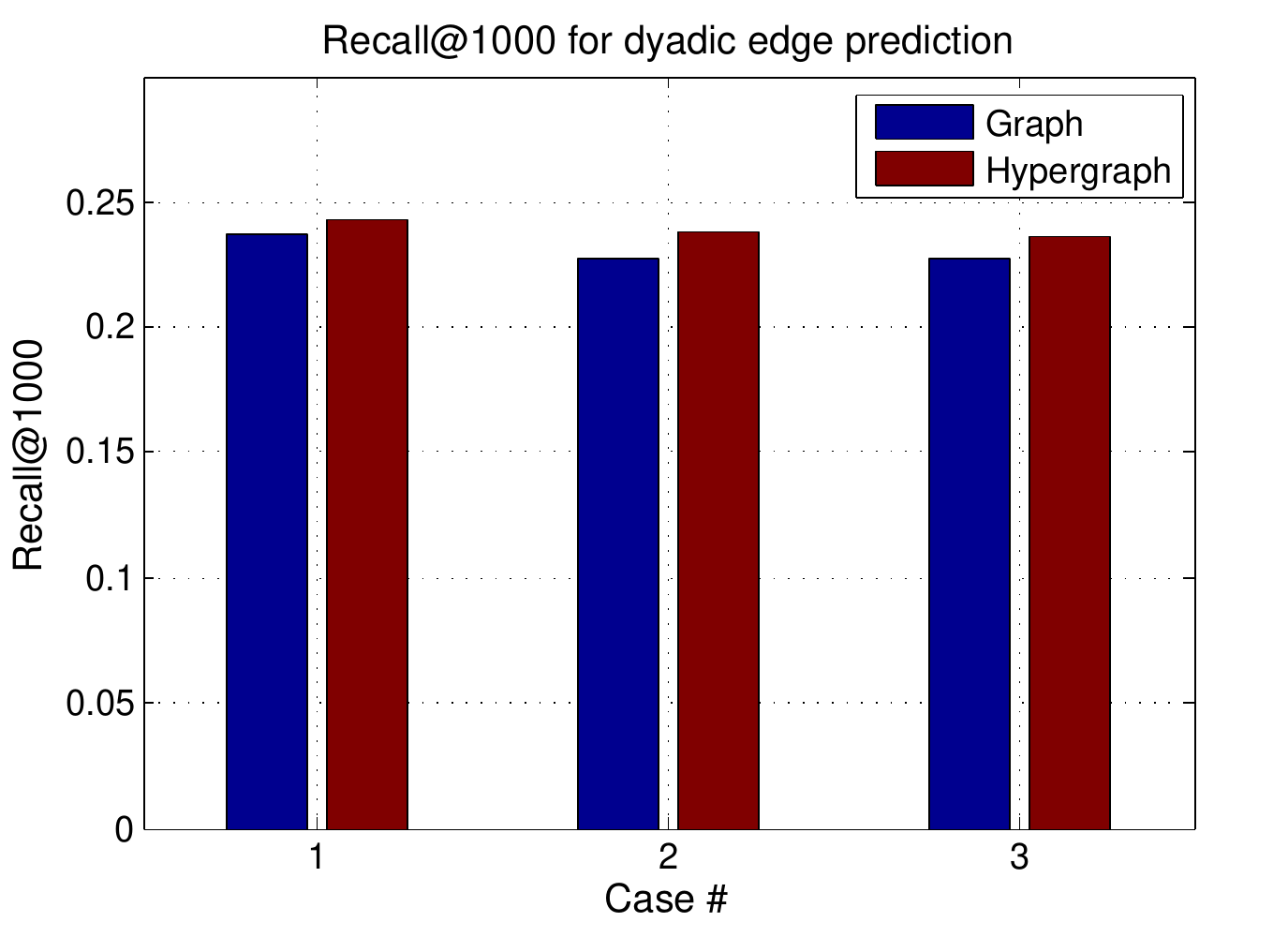}}      
        \caption{Experiment D (Dyadic Link Prediction): (a) AvgPrecision@1000 (b) AvgRecall@1000}
        \label{fig:EXPD}
\vspace{-1.6em}
\end{figure}

This experiment analyzes the trivial case of predicting dyadic links. This experiment consists of three sub-experiments with the following testing and training combinations: A.1-A.5 with training of size $= 5$ years and test period size$ = 4$ years (Case 1); B.1-B.4 with training period size = 4 years and test period size = 3 years (Case 2); and last, training using B.1-B.4 with training period of 3 years and testing using A.1-A.4 with test period size $= 4$ years (Case 3). These cases evaluate the dyadic link prediction under various combination of test and training periods. Results of this experiment are shown in the Figure 9(a),(b). It is clearly visible that the maroon bars (hypergraph) for different sub-experiments (Case 1 to 3) are always aslightly higher than blue bars (graphs). This shows that the performance of graphs and hypergraphs is comparable. Although, graphs are themselves sufficiently capture the information needed to predict dyadic links, the proposed tensor model for hypergraph is robust even to predict dyadic links.


\subsubsection{Discussion}

The above mentioned experiments corroborate about hypergraphs being a better and robust model for higher order collaboration than graphs. Results from these experiments demonstrate higher order collaboration (size$>$4) prediction is very difficult using dyadic graph based approach. Also collaborations of size$>2$ hypergraphs in most cases provide better results with an average increase of approx. 45\% in F-Score for different sizes $\in \{3,4,5,6,7\}$ (Figure 6). In fact our approach is robust for dyadic link prediction as well (Experiment D). Also combined inference from Experiment B and C shows that using the recent (past 3 years) publications we can prediction of a collaboration working together for an extended period of future 5 years. 



%

\section{Conclusion and  Future Work}

In this work we have formulated the problem of higher order collaboration prediction. We show that these problems are much harder than the dyadic edge predictions. We make a pioneering attempt to address the \textit{old edge} prediction problem. For this we propose a novel tensor decomposition based approach. We show that tensors are an excellent way to capture temporal hypergraphs since they perform much better in predicting collaborations of size greater than three (in general higher order hyperedges) in comparison to the dyadic graph representation. Moreover, it also turns out that the hyper-incidence tensor model is robust for dyadic edge prediction as well. In this way we also provide a much needed explicit comparison between graphs and hypergraphs. 

In the future we can work on modifying tensor based approach to address the harder problem of \textit{new link} prediction. We can also use the power of tensors to predict exact future patterns (eg: giving a likelihood of publications $n^{th}$ year in future) by using methods similar to \cite{kolda11}. Another interesting direction is to try out modeling $K$-size collaboration as a $K$-ary Tensors model. 


\section*{Acknowledgment}

The authors would also like to thank Dr. Tamara G. Kolda for her support and also for providing us with the code for Tensor Toolbox \cite{kolda07}. 

\nocite{ex1,ex2}
\bibliographystyle{IEEEtran}
\bibliography{sdmbib}
\end{document}